\documentstyle[prd,aps,preprint]{revtex}
\begin{document}
\draft
\title{$< T^{\mu}_{\nu}>_{ren}$ of the quantized conformal fields 
in the Unruh state in the Schwarzschild spacetime}
\author{Jerzy Matyjasek}
\address{Institute of Physics, Maria Curie Sk\l odowska University\\
pl. Marii Curie Sk\l odowskiej 1, 20-031 Lublin, Poland\\
email: matyjase@tytan.umcs.lublin.pl\\
jurek@iris.umcs.lublin.pl }
\maketitle

\begin{abstract}
The renormalized expectation value of the stress energy  tensor
of the conformally invariant massless fields
in the Unruh state in the Schwarzschild spacetime is constructed.
It is achieved through solving the conservation equation
in conformal space and utilizing the regularity  conditions in a physical
metric.
The relations of obtained results to existing approximations are analysed.

\end{abstract}

\vskip 2cm\noindent{04.62+v, 04.70.Dy \\UMCS-FM-98-16\\August 1998}
\newpage

\section{INTRODUCTION}
%%%%%%%%%%%%%%%%%%%%%%%%%%%

The expectation value of the stress-energy tensor of the conformally
invariant massless  fields in the Unruh state in the Schwarzschild
geometry is known to possess some general features presented for the first
time in the celebrated Christensen and Fulling paper~\cite{Steve}. The asymptotic
behavior of tangential and radial components of $<T_{\nu }^{\mu }>_{ren}$%
and the regularity conditions on the future event horizon are quite
restrictive, and allow construction of a class of approximate tensors.
Further, the numerical data, such as the exact value of the $<T_{\theta
}^{\theta }>_{ren}$ on the future event horizon and the value of the
luminosity may be used in the final determination of the model. It
is very fortunate, that thanks to the excellent numerical analysis carried
out in~\cite{Elster} and~~\cite{Bruce3} we have detailed informations concerning the
overall character of the exact $<T_{\nu }^{\mu }>_{ren}$ of the scalar 
field and consequently the
validity of constructed approximations may by verified. Similar calculations has
been carried out in the case of the  conformal vector field.

The attempts to construct analytical or semianalytical approximations of $%
<T_{\nu }^{\mu }>_{ren}$ in various ``vacuum" states, see Refs. [4-19], 
are motivated, besides self -evident curiosity, by
the fact that they may be used as a source term of the semiclassical
Einstein field equations [20-30] or give rise to the analyses of local and averaged
energetic conditions and quantum inequalities [31-38]. The stress energy tensor is
also useful in the thermodynamic calculations, see for example [20,25,27,29,39,40]. It should be noted that the back 
reaction calculations
are limited to the $< T^{\mu}_{\nu}>_{ren}$ evaluated in the Hartle - Hawking
state.

Recently  such
semianalytic approximations of the stress tensor in the Unruh state have
been presented [16-18]. In Refs. [16,17], where properties of the
one-loop effective action under the conformal transformations have been
used, the starting point is the assumption that in the ultrastatic companion
of the Schwarzschild metric the tangential component of\\
$<\tilde{T}_{\nu}^{\mu }>_{ren}$ has a polynomial form 
\begin{equation}
<T_{\theta }^{\theta }>_{ren}=T p(s)\sum_{n=4}^{N}a_{n}x^{n},
\end{equation}
where $x=2M/r$ and  $T=T_{H}^{4}/90$ $(T_{H} =1/(8 \pi M)).$  The numerical 
factor $p(s)$ depends on the spin of the field
and is given in the Table 1.
Such a choice is in accord with the asymptotic analyses of 
Christensen - Fulling. Further, integrating the conservation
equation in the conformal space and making use of the regularity conditions
in the physical one, a family of the stress tensors could be easily
obtained. 
Taking $ N = 7$ 
as have been done in Refs. [16,17]
leaves two free parameters which  are to be determined
from known value
of the tangential pressure on the event horizon and the luminosity. It
should be noted that it is, in a sense, a minimal choice. On the other
hand, in Ref.~\cite{Visser4} Visser chooses to work in the physical space and uses
a decomposition of the stress tensor allowed by the covariant conservation
equation. The tangential component of $<T_{\nu }^{\mu }>_{ren}$ is also taken
to be of the form (1) with $N\,=\,6,$
this time however in the Schwarzschild spacetime. 
The resulting model has three free parameters which
are determined by performing global unconstrained fit to the numerical data.
The tensors obtained in both models have a similar structure but  differ
in the numerical coefficients. 

In this note we shall show that,
although differently motivated, Visser's model may be obtained
following the steps of Refs. [16,17] with $N = 8.$ The simplest $ N = 5$ 
models proposed by Vaz~\cite{Vaz} and by Barrioz and Vaz~\cite{Vaz1}
has no free parameters, it should be noted however that
the latter uses different decomposition
of the stress tensor in the conformal space. In our constructions 
scalar, spinor, and vector fields are treated simultaneously. 
Presented method (with $n\,=\,1$) may be also employed in 
similar
calculations in Hartle - Hawking
state.

\section{Conformal transformations of the stress - energy tensor}
%%%%%%%%%%%%%%%%%%%%%%%%%%%%%%%%%%%%%%%%%%%%%%%%%%%%%%%%%%%%%%%%%%%%%%%%

Transformational properties of the stress tensor of a conformally coupled
massless field under the scaling transformation $\tilde{g}_{\mu \nu
}\,=\,e^{-2\omega }g_{\mu \nu }$ are described by 
or may be obtained from the 
formulas derived 
by Page~\cite{Page},  Brown and Ottewill~\cite{Brown1}
Brown, Page, and Ottewill~\cite{Brown2}, and Dowker~\cite{Dowker}. 
Under the conformal transformation the one loop effective action of the massless
conformally invariant
fields transforms according to the rule~\cite{Brown1}
%%%%%%%%%%%%%%%%%%%%%%%%%%%%%%%%%%%%%%%%%%%%%%%%%%%%%%%%%%%%%%%%%%%%%%%%%%%
%%%%%%%%%%%%%%%%%%%%%%%%%%%%%%%%%%%%%%%%%%%%%%%%%%%%%%%%%%%%%%%%%%%%%%%%%%%
\begin{equation}
W_{R}[g_{\mu\nu}]\,=\,W_{R}[e^{-2 \omega} g_{\mu\nu}]\,+\,a\, A[\omega;g]\,+\,
b\, B[\omega, g]\,+\,c\, C[\omega, g],
\end{equation}
%%%%%%%%%
where
%%%%%%%%%
\begin{equation}
A[\omega, g]\,=\,\int d^{4}x (- g)^{1/2}\left\{\left( Riem^{2}\,-\,2 Ricc^{2}\,
+\,{1\over 3} R^{2}\right)\omega\,
+\,{2\over 3}\left[ R+3\left(\Box\omega\,-\,\kappa\right)\right]\left(\Box \omega\,
-\,\kappa\right)\right\},
\end{equation}
%%%%%%%%%
\begin{equation}
B[\omega, g]\,=\,\int d^{4}x (-g)^{1/2}\left[ \left( Riem^{2}\,
-\,4 Ricc^{2}\,+\,R^{2}\right)\omega\,
+\,4 R_{\mu\nu}\omega^{;\mu}\omega^{;\nu} -2 R \kappa\,+\,2 \kappa^{2}\,
-\,4\kappa\Box\omega\right],
\end{equation}
%%%%%%%%%%%%
\begin{equation}
C[\omega, g]\,=\,\int d^{4}x (-g)^{1/2}\left\{\left[ R\,+\,3\left(
\Box R\,-\,\kappa\right)\right]\left(\Box R\,-\,\kappa\right)\right\},
\end{equation}
and $\kappa\,=\,\omega_{; \alpha} \omega^{; \alpha}.$
We have distinguished quantities evaluated in the conformal space with a tilde.
The spin-dependent
coefficients $a,\,b\,$ and $c$ are given in Table 1.

Functionally differentiating (2) with respect to the metric tensor
and restricting to the Ricci-flat spaces
one has 
\begin{equation}
< T_{\nu }^{\mu }>_{ren} \,=\,\exp {(-4\omega )}\,{\tilde{T}}_{\nu
}^{\mu }\,+a(s)A_{\nu }^{\mu }\,+\,b(s)B_{\nu }^{\mu }\,+\,c(s) C_{\nu}^{\mu},
\end{equation}
where 
\begin{equation}
A^{\mu \nu }\,=\,8R^{\alpha \mu \nu \beta }\omega _{;\alpha \beta }-{\frac{4%
}{3}}\kappa ^{;\mu \nu }+2g^{\mu \nu }\left( 2\omega ^{;\alpha }\kappa
_{;\alpha }\,+\,\kappa ^{2}\,+\,{\frac{2}{3}}\Box \kappa \right)
\,-\,8\kappa ^{;(\mu }\omega ^{;\nu )}\,-\,8\omega ^{;\mu }\omega ^{;\nu
}\kappa ,
\end{equation}
\begin{eqnarray}
B^{\mu \nu }\, &=&\,8R^{\alpha \mu \nu \beta }\omega _{;\alpha \beta
}\,+\,8R^{\alpha \mu \nu \beta }\omega _{;\alpha }\omega _{;\beta
}\,-8\omega ^{;\mu \alpha }{\omega _{;\alpha }}^{;\nu }\,-\,8\kappa ^{;(\mu
}\omega ^{;\nu )}\,-\,8\kappa \omega ^{;\mu }\omega ^{;\nu }\,  \nonumber \\
&&\quad +\,4g^{\mu \nu }\left( \omega _{;\alpha \beta }\omega ^{;\alpha
\beta }\,+\,\kappa _{;\alpha }\omega ^{;\alpha }\,+\,{\frac{1}{2}}\kappa
^{2}\right) ,
\end{eqnarray}
and
\begin{eqnarray}
C^{\mu\nu}\,=\, g^{\mu\nu}(2\Box\kappa\,+\,3\kappa^2\,+\,6 \omega_{;\alpha}
 \kappa^{;\alpha}) \,-\,12\kappa
 \omega^{;\mu}\omega^{;\nu}\,-\,12\kappa^{;(\mu}\omega^{;\nu)}\,-\,2
 \kappa^{:\mu\nu}.
\end{eqnarray}
The renormalized effective stress tensor has been defined by
\begin{equation}
<T^{\mu \nu}>_{ren}\,=\,2 (-g)^{-1/2}{\delta W_{R}\over \delta g_{\mu \nu}}.
\end{equation}

We observe that since (6) is a general formula its meaning is clear:
the better approximation of the stress tensor in the conformal space is constructed
the better $<T^{\mu}_{\nu}>_{ren}$ in the physical is obtained. There is nothing
less general in working in the conformal space as long as the transformation formulae
are correct.
Moreover, although we heavily used scaling properties of the renormalized stress tensor
the adopted method is neither the Page nor Brown and Ottewill approximation.

Taking the conformal factor in the form $\omega\,=\,1/2 \ln |g_{00}|,$ 
one obtains for $A^{\mu}_{\nu},$ $B^{\mu}_{\nu},$ and $C^{\mu}_{\nu}$ the following
formulae
 \begin{equation}
A^{t}_{t}\,=\,{\frac{{x^6}\,\left( 128 - 240\,x + 113\,{x^2} \right) }
   {128\,{M^4}\,{{\left( 1 - x \right) }^2}}},
\end{equation}
\begin{equation}
A^{r}_{r}\,=\,{\frac{{x^6}\,\left( 32 - 96\,x + 63\,{x^2} \right) }
   {384\,{M^4}\,{{\left( 1 - x \right) }^2}}},
\end{equation}
\begin{equation}
A^{\theta}_{\theta}\,=\,
{\frac{- {x^6}\,\left( 64 - 120\,x + 57\,{x^2} \right)  }
   {384\,{M^4}\,{{\left( 1 - x \right) }^2}}},
\end{equation}

\begin{equation}
B^{t}_{t}\,=\,
{\frac{3\,{x^6}\,\left( 48 - 96\,x + 47\,{x^2} \right) }
   {128\,{M^4}\,{{\left( 1 - x \right) }^2}}},
\end{equation}
\begin{equation}
B^{r}_{r}\,=\,
{\frac{{x^6}\,\left( 16 - 48\,x + 33\,{x^2} \right) }
   {128\,{M^4}\,{{\left( 1 - x \right) }^2}}},
\end{equation}
\begin{equation}
B^{\theta}_{\theta}\,=\,
{\frac{- {x^6}\,\left( 32 - 72\,x + 39\,{x^2} \right)   }
   {128\,{M^4}\,{{\left( 1 - x \right) }^2}}},
\end{equation}

\begin{equation}
C^{t}_{t}\,=\,
{\frac{3\,{x^6}\,\left( 32 - 64\,x + 33\,{x^2} \right) }
   {256\,{M^4}\,{{\left( 1 - x \right) }^2}}},
\end{equation}
\begin{equation}
C^{r}_{r}\,=\,
{\frac{- {{\left( 8 - 9\,x \right) }^2}\,{x^6}  }
   {256\,{M^4}\,{{\left( 1 - x \right) }^2}}},
\end{equation}
\begin{equation}
C^{\theta}_{\theta}\,=\,
{\frac{{x^6}\,\left( 128 - 264\,x + 135\,{x^2} \right) }
   {256\,{M^4}\,{{\left( 1 - x \right) }^2}}}.
\end{equation}
The geometrical terms, i. e. the sum of the last three terms in the 
right hand side of Eq. (6)  have, therefore, a simple form, proportional 
to simple polynomials in $x$ multiplied by the second power of $g_{t t}.$

%%%%%%%%%%%%%%%%%%%%%%%%%%%%%%%%%% Trace anomaly %%%%%%%%%%%%%%%%%%%%%%%%%%%%%%%
It is a well known property of the massless and conformally coupled fields 
that although on the classical level the trace of the stress energy
tensor identically vanish, after quantization it acquires nonzero value.
Making use of the identity 
\begin{equation}
{\delta \over \delta \omega} S[ e^{-2 \omega} g_{\mu \nu}]_{|\omega = 0}\,=\,
-2 g_{\sigma \tau} {\delta \over \delta g_{\sigma \tau}} S[g_{\mu \nu}],
\end{equation}
where $S$ is a functional, one concludes that
the  trace anomaly is given
by a  general formula
\begin{equation}
<T^{\mu}_{\mu}>_{ren}\,=\,a(s) \left({\cal H}\,+\,{2\over 3} \Box R \right)\,+\,
b(s) {\cal G}\,+\,c(s) \Box R,
\end{equation}
where
\begin{equation}
{\cal H}\,=\,Riem^{2}\,-\,2 Ricc^{2}\,+\,{1\over 3} R^{2}
\end{equation}
and
\begin{equation}
{\cal G}\,=\,Riem^{2}\,-\,4 Ricc^{2}\,+\,R^{2}.
\end{equation}
It is a very fortunate coincidence that for the scalar and spinor field
$<\tilde{T}^{\mu}_{\mu}>_{ren}$ in the optical metric is zero.
However, choosing for the vector field the coefficient $c(1)$ 
as predicted by $\zeta-$~function renormalizaton requires
some care. Indeed, in this case the trace of $<\tilde{T}^{\mu}_{\nu}>_{ren}$
does not vanish and, of course, further calculations should reflect this fact.

%%%%%%%%%%%%%%%%%%%%%%%%%%%%%%%%%%%%%%%%%%%%%%%%%%%%%%%%%%%%%%%%%%%%%%%%%%%%%%%%%%%%%%%%
%%%%%%%%%%%%%%%%%%%%%%%%%%%%%%%%%%%%%%%%%%%%%%%%%%%%%%%%%%%%%%%%%%%%%%%%%%%%%%%%%%%%%%%%'
The conservation equation for the line element of the conformal space
\begin{equation}
\tilde{\nabla}_{\mu} <\tilde{T}^{\mu}_{\nu}	>_{ren}\,=\,0,
\end{equation}
reduces to
\begin{equation}
{\partial \over \partial x}<\tilde{T}^{r}_{t}>_{ren}\,-\,{2 (1 - 2 x)\over x (1 - x) }
<\tilde{ T}^{r}_{t}>_{ren}\,=\,0,
\end{equation}
and
\begin{equation}
{\partial\over \partial x} <\tilde{T}^{r}_{r}>_{ren}\,-
\,{2 - 3 x\over x (1 -x )}<\tilde{T}^{r}_{r}>_{ren}\,+
\,{2 -3 x\over x (1 - x)}<\tilde{T}^{\theta}_{\theta}>_{ren}\,=\,0.
\end{equation}

Therefore one concludes that
the stress tensor in the conformal space naturally splits into two parts
\begin{equation}
<\tilde{T}^{\mu}_{\nu}>_{ren}\,=\,{\cal T}^{\mu}_{\nu}\,+\,{9 c \over 8 M^{4}} x^{6} (1 - x)^{2}
\delta^{\mu}_{0}\delta^{0}_{\nu},
\end{equation}
where $ {\cal T}^{\mu}_{\nu} $ is a conserved traceless tensor in the conformal 
space. 
Such a decomposition 
is allowed by the covariant conservation equation in the ultrastatic space and its exact
form is constructed from (21)  evaluated in the optical space. 
Similar term has been introduced into the 
Page approximation by Zannias~\cite{Zannias}. 
It should be noted that in a case of the dimensional renormalization 
the second term in the right hand side of the equation (27) is absent. 

\section{The model}
%%%%%%%%%%%%%%%%%%%%%%%%%%%

As  has been said 
the idea of constructing $<T_{\nu }^{\mu }>_{ren}$ 
in the Schwarzschild spacetime
is to accept (1) and solve
the conservation equation in the ultrastatic space. 
In order to reduce the number of unknown parameters we
transform (1) to the physical space and make use of the regularity condition 
\begin{equation}
\lim_{x \to 1}\,|<T_{\theta }^{\theta }>_{ren}|<\infty .
\end{equation}
Inserting (11 - 19) into (6) with appropriate numerical coefficients 
and making use of (28)
yields
\begin{equation}
a_{7}\,=\,{a\over  48 p t}\,-\,{b\over 32 p t}\,+\,{c \over 32 p t} - 4 a_{4} - 3 a_{5} - 2 a_{6}
\end{equation}
and 
\begin{equation}
a_{8}\,=\,-{7 a \over 384 p t}\,+\,{7 b\over 128 p t}\,+\,{7 c \over 256 t}\,-\,7 + 
3 a_{4} + 2 a_{5} + a_{6},
\end{equation}
where $t\,=\,M^{4} T.$
After substitution of Eqs. (29) and (30) into (1) and performing integration
of the equations (24) and (25) in the conformal
space one has 
\begin{equation}
<\tilde{T}_{t}^{r}>_{ren}=-x^{2}(1-x)^{2}\frac{K}{4M^{4}},
\end{equation}
and  the radial component of 
the stress tensor in the conformal space
%%%%%%%%%%%%%%%%%%%%%%%%%%%%%%%%%%%%%%%%%%%%%%%%%%%%%%%%%%%%%%%%%%%%%%%%%%%%%%%%%%%%%%%%%%
%  radial compt in conf space=T[(1 - d)*x^2 + (-1 + d)*x^3 - x^4*a[4] + 
%  x^5*((2*a[4])/3 - (2*a[5])/3) + 
%  x^8*(21/5 - (9*a[4])/5 - (6*a[5])/5 - (3*a[6])/5) + 
%  x^6*(a[4]/3 + (5*a[5])/12 - a[6]/2) + 
%  x^7*(-16/5 + (9*a[4])/5 + (29*a[5])/20 + (11*a[6])/10)]
%%%%%%%%%%%%%%%%%%%%%%%%%%%%%%%%%%%%%%%%%%%%%%%%%%%%%%%%%%%%%%%%%%%%%%%%%%%%%%%%%%%%%%%%%%
\begin{eqnarray}
<\tilde{T}_{r}^{r}>_{ren}\, &=&\,p\,T \left[\left({a\over 384 p t}\,-\,
{b\over 128 p t}\,+\,{c\over 256 p\,t}\,-\,d \right) x^{2}\,-\,
\left( {a\over 384 p\,t}\,-\,
{b\over 128 p\,t}\,+\,{c\over 256 p\,t}\,-\,d\right) x^{3}\right.\nonumber\\
&&
\,-\,a_{4} x^{4}\,+\,
\left({\frac{2}{3}} a_{4} - {\frac{2}{3}} a_{5}\right) x^{5}
\, +\,\left({\frac{1}{3}} a_{4} + {\frac{5 }{12}} a_{5}- {\frac{1}{2}} a_{6}\right)
x^{6}\,\nonumber\\
&&
+  \left({b\over 40 p\,t}\,-\,{a\over 120 p\,t}\,-\,{c\over 80 p\,t}
-{\frac{16}{5}} + {\frac{9}{5}} a_{4} + {\frac{29 }{20}} a_{5} + {%
\frac{11}{10}} a_{6}\right) x^{7}\nonumber \\
 &&
+\left.\,\left({7\over 640 p\,t}\,-\,{21 b\over 640 p\,t}\,+\,{21 c \over 1280 p\,t}\,+\,
{\frac{21 }{5}} - {\frac{9}{5}} a_{4} - {%
\frac{6 }{5}} a_{5} - {\frac{3}{5}} a_{6}\right) x^{8}\right],
\end{eqnarray}
where $K$ is an integration constant connected to the luminosity and
$d$ is another integration constant. 
Note that the leading behavior of (32) as $x\,\to\,0$ is proportional to $x^{2}$ as
expected.

Now, on general grounds one expects that at large $r$
the leading terms of $- < T^{t}_{t}>_{ren}$ and $< T^{r}_{r}>_{ren}$ should be equal
to $<T^{t}_{r_{*}}>_{ren},$	where $r_{*}$ is the Regge - Wheleer coordinate.
Moreover, the  Christensen - Fulling regularity conditions in the Schwarzschild space,
i. e. conditions for the regularity of the stress - energy tensor on the future event horizon
\begin{equation}
| <T_{v v}>_{ren}|\,<\,\infty ,
\end{equation}
\begin{equation}
| <T^{t}_{t}>_{ren}\,+\,<T^{r}_{r}>_{ren} |\,<\,\infty ,
\end{equation}
and
\begin{equation}
(1 - x)^{-2} |< T_{u u} >_{ren}| \,<\,\infty
\end{equation}
as $x \to 1, $
where
\begin{equation}
< T_{u u}>_{ren} \,=\,{1\over 4}\left[ (1-x) (<T^{r}_{r}>_{ren}\,-\,
<T^{t}_{t}>_{ren})\,+\,{2 K\over M r^2}\right],
\end{equation}
and
\begin{equation}
< T_{v v}>_{ren}\,=\,<T_{u u}>_{ren}\,-\,{K\over M^{2} r^{2}}
\end{equation}
allows to reduce the number of unknown
parameters to three. 
This conditions together with (28) ensure that the stress-energy tensor measured in the 
local frames on the future event horizon will be finite.

The conditions (28) and (34) are already satisfied while the remaining ones after
simple calculations give 
%%%%%%%%%%%%%%%%%%%%%%%%%%%%%%%%%%%%%%%%%%%%%%%%%%%%%%%%%%%%%%%%%%%%%%%%%%%%%%%%%%%%%%%%%%%
% d =  + (7*a[4])/30 + (17*a[5])/120 + a[6]/20
%
%
% K = T ((112*M^4)/5 - (14*M^4*a[4])/15 - (17*M^4*a[5])/30 - (M^4*a[6])/5)
%%%%%%%%%%%%%%%%%%%%%%%%%%%%%%%%%%%%%%%%%%%%%%%%%%%%%%%%%%%%%%%%%%%%%%%%%%%%%%%%%%%%%%%%%%%
\begin{equation}
K\,=\,p\,t \left(-\,{\frac{14}{15}} a_{4}\,\,-\,{\frac{17}{30%
}} a_{5}\,-\,{\frac{1}{6}} a_{6}\right)\,+\,{11 a\over 60}\,+\,{b\over 5}\,-\,{c\over 10},
\end{equation}
and 
\begin{equation}
d\,=\,{\frac{7}{30}} a_{4}\,+\,{\frac{17}{120}}
a_{5}\,+\,{\frac{1}{20}} a_{6}\,-\,{83 a\over 1920 p\,t}\,-\,{37 b\over 640 p\,t}\,+
\,{37 c\over 1280 p\,t}.
\end{equation}
%%%%%%%%%%%%%************************************************************
Returning to the physical space, after some algebra, one obtains the mean value of the
energy momentum tensor of the quantized, massless, conformally invariant field in 
the Unruh state
\begin{eqnarray}
<T^{t}_{t}>_{ren}\,&=&
\,{p T\over 1 - x}\left\{ 
\left({c\over 40 p \,t}\,-\,{11 a\over 240 p\, t}\,-\,{b\over 20 p\, t}
\,+\,{7\over 30} a_{4}\,+\,{17\over 120} a_{5}\,+\,{a_{6}\over 20}\right) x^{2}
- a_{4} x^{4}\right.\nonumber\\
&&
,-\,\left({5\over 3} a_{4} + {4\over 3} a_{5}\right) x^{5}\,+ \,
\left(  {a\over p\,t}\,+\,{9 b \over 8 p\,t}\,-\,{3 c\over 4 p\,t}\,-\,2 a_{4}\,-\,
{7\over 4} a_{5}\,-\,{3\over 2} a_{6}\right) x^{6}\nonumber\\
&&\left.+\,
\left( {7 c \over 10 p\,t}\,-\,{109 a\over 120 p\,t}\,-\,{41 b\over 40 p\,t}\,+\,{21\over 5} a_{4}\,+\,
{14\over 5} a_{5}\,+\,{7\over 5} a_{6}\right) x^{7}\right\}
\end{eqnarray}

 \begin{eqnarray}
<T^{r}_{r}>_{ren}\,&=&
\,{p T\over 1 - x}\left\{ 
\left({11 a\over 240 p\,t}\,+\,{b\over 20 p\,t}\,-\,{c\over 40 p\,t}
\,-\,{7\over 30} a_{4}\,-\,{17\over 120} a_{5}\,-\,{a_{6}\over 20}\right) x^{2}
- a_{4} x^{4}\right.\nonumber\\
&&
-\,\left({1\over 3} a_{4} + {2\over 3} a_{5}\right) x^{5}\,+ \,
\left(  {a\over 12 p\,t}\,+\,{b \over 8 p\,t}\,-\,{c\over 4 p \,}\,-\, 
{1\over 4}a_{5}\,-\,
\,-\,{1\over 2} a_{6}\right) x^{6}\nonumber\\
&&+\,\left.
\left( {3 c \over 10 p\,t}\,-\,{9 a\over 40 p\,t}\,-\,{9 b\over 40 p \,t}\,+\,{9\over 5} a_{4}\,+\,
{6\over 5} a_{5}\,+\,{3\over 5} a_{6}\right) x^{7}\right\}
\end{eqnarray}

\begin{equation}
< T^{\theta}_{\theta}>_{ren}\,=
\,p T \left\{ a_{4} x^{4}\,+\,\left(2 a_{4}\,+\,a_{5}\right) x^{5}\,+\,
\left({c\over 4 p\,t}\,-\,{a\over 6 p\,t}\,-\,{b\over 4 p\,t}\,+\,3 a_{4}\,+\,2 a_{5}\,+\,a_{6}
\right) x^{6}\right\}
\end{equation}
and
\begin{equation}
<T^{r}_{t}>_{ren}\,=\,- x^{2}{K\over 4 M^{4}},
\end{equation}
where $K$ is given by
\begin{equation}
K\,=\,{1\over 60}\left( 11 a\,+\,12 b\,-\,6 c\right)\,-\,{p T\over 30}\left(28 a_{4}\,
+\,17 a_{5}\,-6 a_{6}\right).
\end{equation}
Generalizations to greater $N$ are obvious, however, it seems that 
the more complicated formulae
are of little use. It should stressed that, by construction, obtained 
tensors satisfy all regularity
and consistency requirements.

\section{Discussion}
%%%%%%%%%%%%%%%%%%%%%%%%%%%

In order to compare just obtained  $<T^{\mu}_{\nu}>_{ren}$
with those constructed by Visser in Ref.~\cite{Visser4}
first we introduce  a new set of unknown parameters $k_{i}$ defined as 
\begin{equation}
a_{4}\,=\,k_{4},
\end{equation}
\begin{equation}
a_{5}\,=\, -2 k_{4} + k_{5},
\end{equation}
and 
\begin{equation}
a_{6}\,=\,k_{4} -2 k_{5} + k_{6}  + {1\over p \,t}\left( {a\over 6} + {b\over 4} -
{c \over 2}\right).
\end{equation}
In terms of $k_{i}$ the stress tensor becomes
%%%%%%%%%%%%%%%%%%%%%%%%%%%%%%%%%%%%%%%%%%%%%%%%%%%%%%%%%%%%%%%%%%%%%%%%%%%%%%%%%%%
% Tensortt = T*x^2/(1-x) (-24/5 + k5/24 + k6/20 - k4*x^2 + (k4 - (4*k5)/3)*x^3 + 
%  (96 + (5*k5)/4 - (3*k6)/2)*x^4 + (-432/5 + (7*k6)/5)*x^5)
%
%=============================================================================
% Tensorrr = T*x^2/(1-x) (24/5 - k5/24 - k6/20 - k4*x^2 + (k4 - (2*k5)/3)*x^3 +
% ((3*k5)/4 - k6/2)*x^4 + 
%  (-48/5 + (3*k6)/5)*x^5)
%=============================================================================
%
% Tensor_ang  = T*x^4*(k4 + x*(k5 + k6*x))
%=============================================================================
%%%%%%%%%%%%%%%%%%%%%%%%%%%%%%%%%%%%%%%%%%%%%%%%%%%%%%%%%%%%%%%%%%%%%%%%%%%%%%%%%%%%%
\begin{eqnarray}
<T^{t}_{t}>_{ren}\,=&&\,{\frac{p\,T x^{2}}{1- x}}\left\{ {\frac{k_{5}}{24}} + {%
\frac{k_{6}}{20}} - {3\over 80\,p\,t}\left( a\,+\, b\right)\,- \,k_{4} x^{2}\,+
\,\left(k_{4} -{\frac{4}{3}}
k_{5}\right) x^{3}\,+\right.  \nonumber \\
&&\left.\left[{3\over 4\,p\,t}\left( a + b\right) + {\frac{5}{4}} k_{5} - 
{\frac{3}{2}} k_{6}\right] x^{4} \,+\,\left[{\frac{7%
}{5}} k_{6} - {\frac{27}{40\,p\,t}}\left(a + b\right)\right] x^{5}\right\},
\end{eqnarray}
\begin{eqnarray}
<T^{r}_{r}>_{ren}\,=&&\,
{\frac{p\,T x^{2}}{1- x}}\left\{ {3\over 80\,p\,t}\left( a\,+\, b\right) - {\frac{%
k_{5}}{24}} - {\frac{k_{6}}{20}} \,-\,k_{4} x^{2}\,+\,\left( k_{4} - {\frac{2}{3}}
k_{5}\right)x^{3}\,+\,\right.  \nonumber \\
&&\left.\left({\frac{3}{4}} k_{5} - {\frac{k_{6}}{2}}\right) x^{4}\,+\,\left[{\frac{3}{5}}
k_{6} - {3\over 40\,p\,t}\left(a + b\right)\right] x^{5}\right\},
\end{eqnarray}
and 
\begin{equation}
<T^{\theta}_{\theta}>_{ren}\,=\,p\,T x^{4} (k_{4} + k_{5} x + k_{6}
x^{2}).
\end{equation}
The net flux is described by (43), where
the integration constant $K$ is expressed in terms of  $k_{i}$ as
\begin{equation}
K\,=\,{1\over 60}\left[9 a + 9 b - 2 \left( 5 k_{5} + 6 k_{6}\right) p t\right].
\end{equation}
Eqs(48 - 51) are equivalents of the $<T^{\mu}_{\nu}>_{ren}$ constructed recently 
by Visser~\cite{Visser4}. 

The parameters $a_{4},\,a_{5},\,a_{6}$ or equivalently $k_{4},\,k_{5},\,k_{6}$
are to be determined from the available numerical data. Such a procedure 
has proven to be very useful
in constructing highly accurate analytical approximations to the exact stress tensor.
In it simplest form one needs the horizon value of, say, $< T^{\theta}_{\theta}>_{ren}$
the luminosity and one additional piece of information. When the results of detailed numerical
calculations are known
one may perform  unconstrained fit to the totality of the available data.  In this approach
known value of the luminosity is used by (51) as a consistency check rather than an input.
This procedure has been adopted recently by Visser in the case of the scalar field.

Unfortunately, detailed calculations, even if executed, are rarely published,
rather, the overall character of the stress tensor is presented graphically.
However, in the static and spherically - symmetric geometries it is relatively
easy to construct the asymptotic characteristics of $<T^{\mu}_{\nu}>_{ren}.$
Consequently, with $<T^{\theta}_{\theta}(1)>_{ren}$ and the luminosity treated as input
one may construct reasonable approximation.
Indeed, taking into account a more restrictive hypothesis $N = 7,$ one concludes
that the additional constraint 
\begin{equation}
3 a_{4} + 2 a_{5} + a_{6} - {1\over 384\,p\,t}\left(7 a + 21 b + 
{21\over 2} c\right)\,=\,0,
\end{equation}
results, after some rearrangement in the stress tensor
\begin{eqnarray}
 \nonumber
 <T^{t}_{t}>_{ren}\,&=&\,
 - {{x^2}\over 4 M^4 (1 - x)}\left[\, K + \left( 4\,h\, + 24\,K\, 
 -\,{119\over 32}a\,-\,{123\over 32}\,+\,{27\over 64}c\right){x^2} \right.
 \nonumber\\
 &-& 
 \left(4\,h\,+\, 56\,K\,\,-\,{303\over 32}a\,-\,{339\over32}b\,+\,
 {234\over 64}c\right){x^3} \nonumber\\
 &&       
 \left.+\,\left( 30\,K\, -\,{297\over 32}a\,-\,{357\over 32}b\,+\,{405\over 64}c\right) x^{4}
 + \left({113\over 32}a\,+\,{141\over 32}b\,-\,{189\over 64}c\right){x^5}\right],  
 \end{eqnarray}
 \begin{eqnarray}
 \nonumber
 <T^{r}_{r}>_{ren}\,&=&\,
 {x^2\over 4 M^4 (1 - x)}\,\left[ K - \left(4\,h\,+ 24\,K\, -\,{119\over 32}a\,-\,
 {123\over 32}b\,+\,{27\over 64}c\right){x^2}\right. \\
 && + 
 \left(4\,h\, + 40\,K\,-\,{211\over 32}a\,-\,{321\over 32}b\,+\,
 {135\over 64}c\right) x^{3}\nonumber\\
 &&
  \left. -\left( 18\,K\,-{113\over 32}a\,-\,{141\over 32}b\,+\,
 {189\over 64}c\right)x^{4} \,-\,
 \left( {21\over 32}a\,+\,{33\over 32}b\,-\,{81\over 64}c\right)x^{5}\right], 
 \end{eqnarray}
 and
 \begin{eqnarray}
 \nonumber
 < T^{\theta}_{\theta}>_{ren}\,&=&\,
 {x^{4}\over M^{4}}\left[ h\,+\,6 K\,-\,{119\over 128}a\,-\,{123\over 128}b\,-\,
 {27\over 256}c\,-\,\left( 6 K\,-\,{69\over 64} a\,-\,{81\over 64}b\,+\,{81\over 128}c
 \right) x\right.\nonumber\\
 &&
 \left. \,-\,\left( {19\over 128}a\,+\,{39\over 128}b\,-\,{135\over 256}c\right) x^{2}
 \right]
 \end{eqnarray}
 where $h\,=\, M^{4} \langle T^{\theta}_{\theta}(2 M)\rangle.$
It is interesting to note that both $N = 8$ and $N = 7$ hypotheses yield 
the same structure of the renormalized stress tensor in the Unruh state.
Substituting the spin-dependent coefficients taken from table 1 for scalar
and vector fields one easily obtains results presented in Refs.~\cite{J3} and~\cite{J4}.
In a case of the conformal vector field the coefficient $c(1)$ has been taken 
as predicted by $\zeta -$function renormalization. Therefore, 
one can draw a conclusion that from the point of view of the
applied method the only difference between the results of [16,17] and [18]
is the choice of $N$ in (1). On the other hand, Barrioz and Vaz~\cite{Vaz1}
take $N=5$ ( i. e. there are no free parameters left) 
and  use more complicated decomposition
of $<\tilde{T}^{\mu}_{\nu}>_{ren}$ in the optical space
\begin{equation}
<\tilde{T}^{\mu}_{\nu}>_{ren}\,=\,{\cal T}^{\mu}_{\nu}\,+\,\left(\alpha\, \delta^{\mu}_{0}
\delta^{0}_{\nu}\,+\,\beta\, \delta^{\mu}_{\nu}\right)< {\tilde T}^{\sigma}_{\sigma}>_{ren},
\end{equation}
where $\alpha$ and $\beta$ are coefficients subjected to the obvious condition
$\alpha\,+\,4 \beta\,=\,1.$

 \section{Concluding remarks}
 %%%%%%%%%%%%%%%%%%%%%%%%%%%%%%%%%
 
In this work our goal was to 
construct $< T^{\mu}_{\nu}>_{ren}$ in the Unruh state and to
investigate how the choice of $N$ in Eq. (1)
affects the resulting stress-energy tensor. Although
our analyses have been limited to $N \leq 8$ 
it seems that a three parameter family of the stress tensor is of sufficient generality.
Since apparently the ambitious 
plan to construct the approximate stress tensor in the Unruh state using the polynomial 
ansatz and appropriate regularity conditions as the only  available {\it a priori}
informations has failed
it seems that the  presented method (or the methods closely related)
are the only one which
would give analytical formulae able to reconstruct the exact 
$< T^{\mu}_{\nu}>_{ren}$
to a high accuracy. Moreover, the price one should pay for such quality of the
approximation
is rather small: just two or three pieces of numeric data, 
As far as we know there are neither numerical calculations 
nor asymptotic analyses concerning the vacuum polarization 
effect of the conformally coupled
massless spinor field in the Schwarzschild spacetime and consequently
the stress-energy tensor cannot be determined completely.
We expect however,
that the general formulae supplemented by additional pieces  of
numerical data would 
give a good approximation of the exact stress tensor in this case also.

Finally, we remark that a similar method, with different asymptotics 
may be used in construction
of $<T^{\mu}_{\nu}>_{ren}$ in the Hartle - Hawking state, specifically,
the results of Refs.~\cite{J} may be rederived~\cite{J7}.
We intend to 
return to this group of problems in a separate publication.

\begin{table}
\caption{The coefficients $a(s),$ $b(s),$ $c(s),$ and $p(s)$ for fields of helicity s.
Fos $s = 1$ the coefficient $c$ is nonzero as predicted by point separation
and  $\zeta-$function renormalization. 
Dimensional regularization
gives $ c(1) = 0.$}
\begin{tabular}{dddd}
$s$ &$ 0$ &$ 1/2$ &$ 1$ \\ 
$5760\pi ^{2}a(s)$ &$ 3$ &$ 9$ &$ 36$ \\ 
$5760\pi ^{2}b(s)$ &$ -1$ &$ -11/2$ &$ -62$ \\ 
$5760\pi ^{2}c(s)$ &$ 0$ &$ 0$ &$ -60$ \\
$p(s)$ & $1$ & $7/4$ & $2$\\
\end{tabular}
\label{table1}
\end{table}

%%%%%%%%%%%%%%%%%%%%%%%%%%%%%%%%%%%%%%%%%%%%%%%%%%%%%%%%%%%%%%%%%%%%%%%%%%%%%%%%%%%%%%%%%%
%%%%%%%%%%%%%%%%%%%%%%%%%%%%%%%%%%%%%%%%%%%%%%%%%%%%%%%%%%%%%%%%%%%%%%%%%%%%%%%%%%%%%%%%%%

\end{document}